\documentclass{webofc}
\usepackage[varg]{txfonts}   
\usepackage{caption}
\usepackage{subcaption}
\usepackage{hyperref}
\begin{document}
\title{Distributed training and scalability for the particle clustering method UCluster}
%
%

\author{\firstname{Olga} \lastname{Sunneborn Gudnadottir}\inst{1}\fnsep\thanks{\email{olga.sunneborn.gudnadottir@cern.ch}} \and
        \firstname{Daniel} \lastname{Gedon}\inst{2} \and
        \firstname{Colin} \lastname{Desmarais}\inst{4}\and
        \firstname{Karl} \lastname{Bengtsson Bernander}\inst{3}\and
        \firstname{Raazesh} \lastname{Sainudiin}\inst{4, 5}\and
        \firstname{Rebeca} \lastname{Gonzalez Suarez}\inst{1}
}

\institute{Department of Physics and Astronomy, Division of High Energy Physics, Uppsala University 
\and      Department of Information Technology, Division of Systems and Control, Uppsala University
\and      Department of Information Technology, Division of Visual Information \& Interaction, Uppsala University
\and      Department of Mathematics, Uppsala University
\and      Combient Competence Centre for Data Engineering Sciences, Uppsala University
          }

\abstract{%
    In recent years, machine-learning methods have become increasingly important for the experiments at the Large Hadron Collider (LHC). They are utilised in everything from trigger systems to reconstruction and data analysis. The recent UCluster method is a general model providing unsupervised clustering of particle physics data, that can be easily modified to provide solutions for a variety of different decision problems.  In the current paper, we improve on the UCluster method by adding the option of training the model in a scalable and distributed fashion, and thereby extending its utility to learn from arbitrarily large data sets. UCluster combines a graph-based neural network called ABCnet with a clustering step, using a combined loss function in the training phase. The original code is publicly available in TensorFlow~v1.14 and has previously been trained on a single GPU. It shows a clustering accuracy of 81\% when applied to the problem of multi-class classification of simulated jet events. Our implementation adds the distributed training functionality by utilising the Horovod distributed training framework, which necessitated a migration of the code to TensorFlow~v2. Together with using parquet files for splitting data up between different compute nodes, the distributed training makes the model scalable to any amount of input data, something that will be essential for use with real LHC data sets. We find that the model is well suited for distributed training, with the training time decreasing in direct relation to the number of GPU's used. However, further improvements by a more exhaustive and  possibly distributed hyper-parameter search is required in order to achieve the reported accuracy of the original UCluster method.
}
\maketitle
\section{Introduction}
\label{intro}
Although machine-learning methods have been used in high energy physics for more than 50 years, recent years have seen a substantial increase in their variety and prevalence (see e.g.~\cite{rev}). This can be connected to different factors, such as, the increased need for more precise methods in experiments, the fast-paced development of novel machine-learning methods coming from both within the academic setting and the private sector, and improvements in computer hardware leading to a larger computing capacity. 

Within the Large Hadron Collider (LHC) experiments, several research problems are already solved with machine-learning; neural networks and boosted decision trees are used for e.g.~flavour tagging jets, separating signal from background in analysis, and particle identification~\cite{elias}. Improving these methods, extending their reach and developing new ones is currently an active field of study. In addition to the above, machine-learning has been proposed to solve such diverse problems as collision and detector simulation, trigger decision-making, model-independent searches for new physics, jet substructure studies and much more~\cite{hepmllivingreview}.

A lot of the development in machine-learning methods for High Energy Physics is done on data sets that have already undergone a great deal of event selection, and are small enough to allow for training on a single Graphical Processing Unit (GPU) used as the compute node. For more complex models, such as graph networks, generative adversarial networks, or any large enough model, this type of training is limited by its lack of scalability, i.e.~its capacity to handle growing amounts of data. This problem can be ameliorated by distributing the training over multiple workers in a cluster of several compute nodes, and thereby increasing the capacity for data ingestion by utilising the distinct storage devices and the working parallel random access memory of several GPU nodes. Using distributed training algorithms also reduces the overall training time by effectively increasing the batch size, which in turn mitigates the problem of prohibitively long training times. In this paper, we apply distributed training to the recently proposed UCluster method~\cite{Mikuni:2020qds} for unsupervised clustering of particle physics data, with the goal of both speeding up the training and making it scalable to arbitrarily large data sets.

UCluster is a neural network for unsupervised clustering on particle collider data. It creates a latent space using a classification network and then clusters particles, that are close in the latent space, together. 

Depending on the desired properties of the clusters, different classification objectives can be used to create different latent spaces. By choosing jet mass classification, the model produces clusters of jets with the same mass, by choosing event classification the model produces clusters of events with similar properties, etc. This makes UCluster general and highly adaptable, and it has the potential to be useful for several physics problems relating to particle or event classification. So far, it has shown promising results when applied to multi-class classification of jets and anomaly detection at the event level. However, with all training and evaluation executed on a single GPU, the size of the input data is limited to the number of events that can be loaded onto one GPU memory simultaneously. This excludes modifying UCluster, as implemented in a single machine setup~\cite{Mikuni:2020qds}, to any task that requires training on bigger sample sizes, e.g.~full data samples from the LHC experiments.

Unsupervised multi-class classification is something that could be of interest in precision measurements, e.g.~in cases where simulated data are not precise enough to be used for background estimation. In this case, data-driven methods, i.e.~methods that use real data to estimate the background, are commonly used. This approach can quickly become involved if it has to be done for more than one background process. Instead, an unsupervised multi-class classification method could be applied directly to data, labelling the processes without the need for multiple background fits. UCluster is reported to have an 81\% clustering accuracy when applied to the problem of classifying the fat jets of the HLS4ML LHC Jet data set~\cite{multiclassdata} into clusters in which the member jets all originate from the same particle, making use of particle mass classification. However, for data-driven background estimation, the data sets could become several orders of magnitudes bigger. 

Anomaly detection is interesting in particle physics, since it is model-independent by nature and can be used to find deviations from the Standard Model, which can then be used as the basis for new studies. UCluster has been applied to the R\&D Dataset for the LHC Olympics 2020 Anomaly Detection Challenge~\cite{gregor_kasieczka_2019_2629073} and reports an increase in signal-to-background ratio from 1\% to 28\%, in which the signal represents the anomalies. This is accomplished through all anomalies ending up in the same cluster. Making this setup scalable would open up the possibilities of looking at larger data sets or even the full experimental data samples of the large LHC collaborations, as has been proposed in reference~\cite{mining}.

To achieve distributed and scalable training, we made use of Apache Spark \cite{spark}, an open-source distributed general-purpose cluster-computing framework, which creates an architecture over a cluster of multiple compute nodes for distributed data processing through a distributed file system, that splits and stores the data for processing in a fault-tolerant manner.
We set this up on the Databricks~\cite{databricks} platform, which allows for easy creation of Apache Spark clusters. This setup bypasses many of the challenges of processing large data sets such as cluster management, unreliable hardware or running out of memory on a single GPU. We used the distributed deep learning framework Horovod~\cite{horovod}, which contains a wrapper to run distributed training in Spark clusters, to run the training. Once a training algorithm has been set up with Horovod, it can be run on any number of GPUs (including only one) without any changes in the code. Distributing the training across multiple GPUs means an effective increase in the batch size of the underlying stochastic gradient descent optimisation algorithm, leading to a faster convergence of the optimisation functions. We expect the training time to be in inverse proportion to the number of GPUs. 

\section{Model and data}
The details of the UCluster model can be found in the original reference~\cite{Mikuni:2020qds}, but some elements needed to understand the rest of this paper will be repeated here. Consider a data set from a particle collider, which has already gone through digitisation and object reconstruction. The reconstructed objects are then represented as nodes in the graph-based neural network known as ABCnet~\cite{ABC}, a classification net that aids in the over-arching classification problem.

This classification net needs to be optimised to create a suitable latent space on the overarching clustering problem of the UCluster model, i.e.~a space in which particles with similar properties are close to each other, making clustering possible. 
The ABCnet is pre-trained for a number of epochs and then a classical k-means algorithm is applied to the latent space. The resulting clusters are used to initialise the cluster centroids in the full model, which is a Deep k-means algorithm that combines the classification net with clustering~\cite{deepk}. 
The full model is trained end-to-end, with the combined classification and clustering loss, and the trained model assigns every data-point to a cluster. The code is written in TensorFlow v.1~\cite{tensorflow}.

In this paper, we will mimic the first use case demonstrated in the original paper -- multi-class classification of fat jets from the HLS4ML LHC Jets data set -- using distributed training. 
The classification objective is mass classification of jets. 
The goal is to produce three clusters of jets, each of which contains only jets originating from either a W-boson, Z-boson or top quark. 

This HLS4ML data set contains high p\(_\text{T}\) jets originating from W-bosons, Z-bosons and top-quarks from simulations of LHC proton-proton collisions at \(\sqrt{\text{s}}=13\) TeV, which have subsequently been run through a parametric description of a generic LHC detector. They have then been reconstructed using the anti-k\(_\text{T}\) algorithm \cite{antikt} with radius parameter R\(=0.8\). The data consist of various high-level features of each of up to 100 constituent particles of each jet. Table~\ref{tab-1} shows the data features of each constituent used for this project.
\begin{table}
\centering
\caption{Data features used as input in the UCluster method.}
\label{tab-1}       
\begin{tabular}{lll}
\hline
Variable & Description  \\\hline
\(\Delta\eta\) & Difference in pseudorapidity \(\eta\) between jet constituent and jet.\\
\(\Delta \phi\) & Difference in azimuthal angle \(\phi\) between jet constituent and jet. \\
\(\log(\text{p}_\text{T})\) & Logarithm of the constituent transverse momentum \(\text{p}_\text{T}\). \\
\(\log(\text{E})\) & Logarithm of the constituent energy E. \\
\(\log(\frac{\text{p}_\text{T}}{\text{p}_\text{T}\text{(jet)}})\) & Logarithm of the constituent transverse momentum relative to the \\
&jet transverse momentum.\\
\(\log(\frac{\text{E}}{\text{E}\text{(jet)}})\) & Logarithm of the constituent energy relative to the jet energy. \\
\(\Delta R\)& Distance defined as \(\sqrt{\Delta\eta^2+\Delta\phi^2}\) \\
PID & Particle ID \cite{PDG} \\\hline
\end{tabular}
\end{table}
 
The data are stored in two HDF5 files~\cite{hdf5}: one training data set and one validation data set. Following the unconventional nomenclature of the HLS4ML challenge~\cite{multiclassdata}, we call the two data sets, used for development, the~\textit{training} and~\textit{testing} data sets and the data set we test our final model on as the~\textit{validation} data set.

\section{Modifying the code for scalability and distributed training}
To make use of the Horovod framework, we needed to migrate the code to Tensorflow v2. We then extended the model to allow for distributed training. We thus have three incarnations of the code:\\
\textbf{The original code} which is obtained from the GitHub of the original authors \cite{uclustergit}. This was used to validate our setup against.\\
\textbf{The original code migrated to TensorFlow~v2}. This was done with an automated function supplied by TensorFlow and relies on TensorFlow~1 compatible functions. This code was validated against the original code and shows comparable results in both training behaviours, results and execution time. \\
\textbf{The distributed model}. This is described in the next section.

\subsection{Distributed training}
With the Tensorflow~v2 code as a starting point, we used the HorovodRunner, the Horovod Databricks Application Programming Interface (API), to be able to run distributed training. This was done by creating a HorovodRunner instance and passing it the training function. The training function had to be modified for use with Horovod, including changes, such as, increasing the learning rate to compensate for the bigger effective batch size and handling of checkpoints to ensure consistent saving to and initialisation from them. With our data on the local driver, we initialise the HorovodRunner instance, which copies the data to each GPU. The Horovod framework takes care of the distributed training using a distributed optimiser to accumulate gradients across multiple GPUs, and saves model checkpoints at regular intervals. The learning rate is scaled by the number of GPUs used.  Currently, all data is copied onto each GPU, which limits the scalability to what can be fit into a single GPU memory. This will be addressed by copying only fractions of data to each GPU at a time.

\subsection{Scalability}
The UCluster repository includes a pre-processing script specifically for the HLS4ML data set which we use to extract the relevant features from it (see Table~\ref{tab-1}). After pre-processing, the data are contained in a single file and will have to be loaded in their entirety onto the local driver before training. This currently puts limits on the scalability of our setup, and will be addressed in future work by writing data loaders directly into the distributed file system.

\section{Training and evaluation}
\begin{figure}[h]
    \centering
    \begin{subfigure}[b]{0.47\textwidth}
        \centering
        \includegraphics[width=\textwidth]{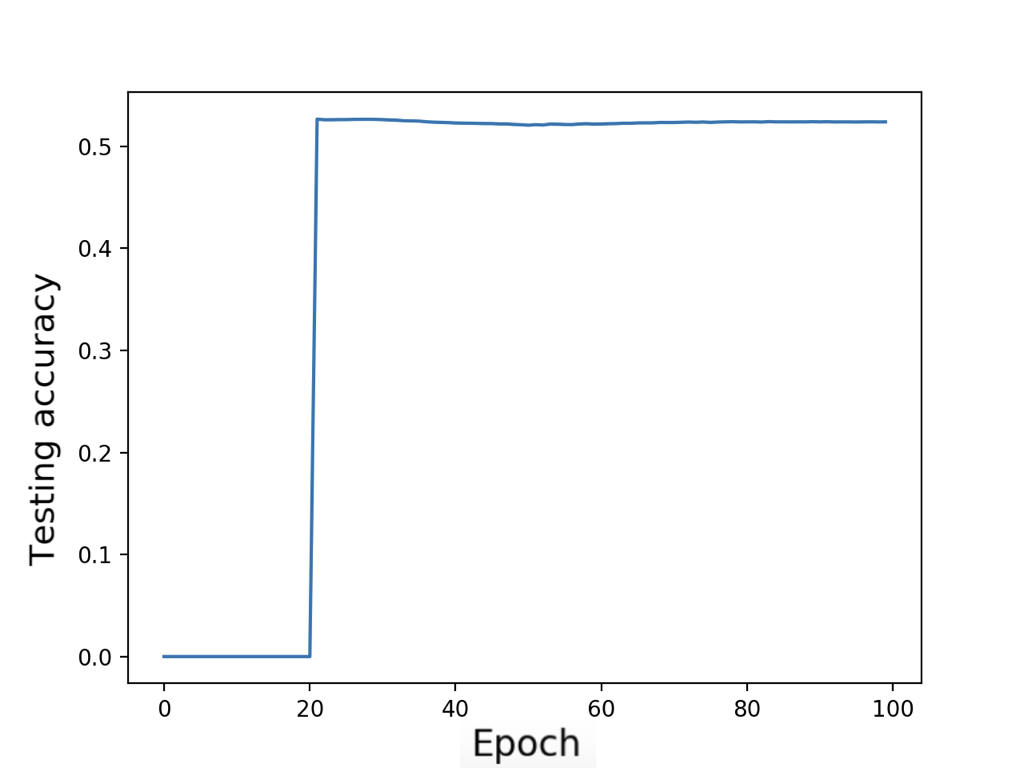}
        \caption{Testing accuracy of single machine training with original code.}
        \label{v1}
    \end{subfigure}
    \hfill
    \begin{subfigure}[b]{0.47\textwidth}
        \centering
        \includegraphics[width=\textwidth]{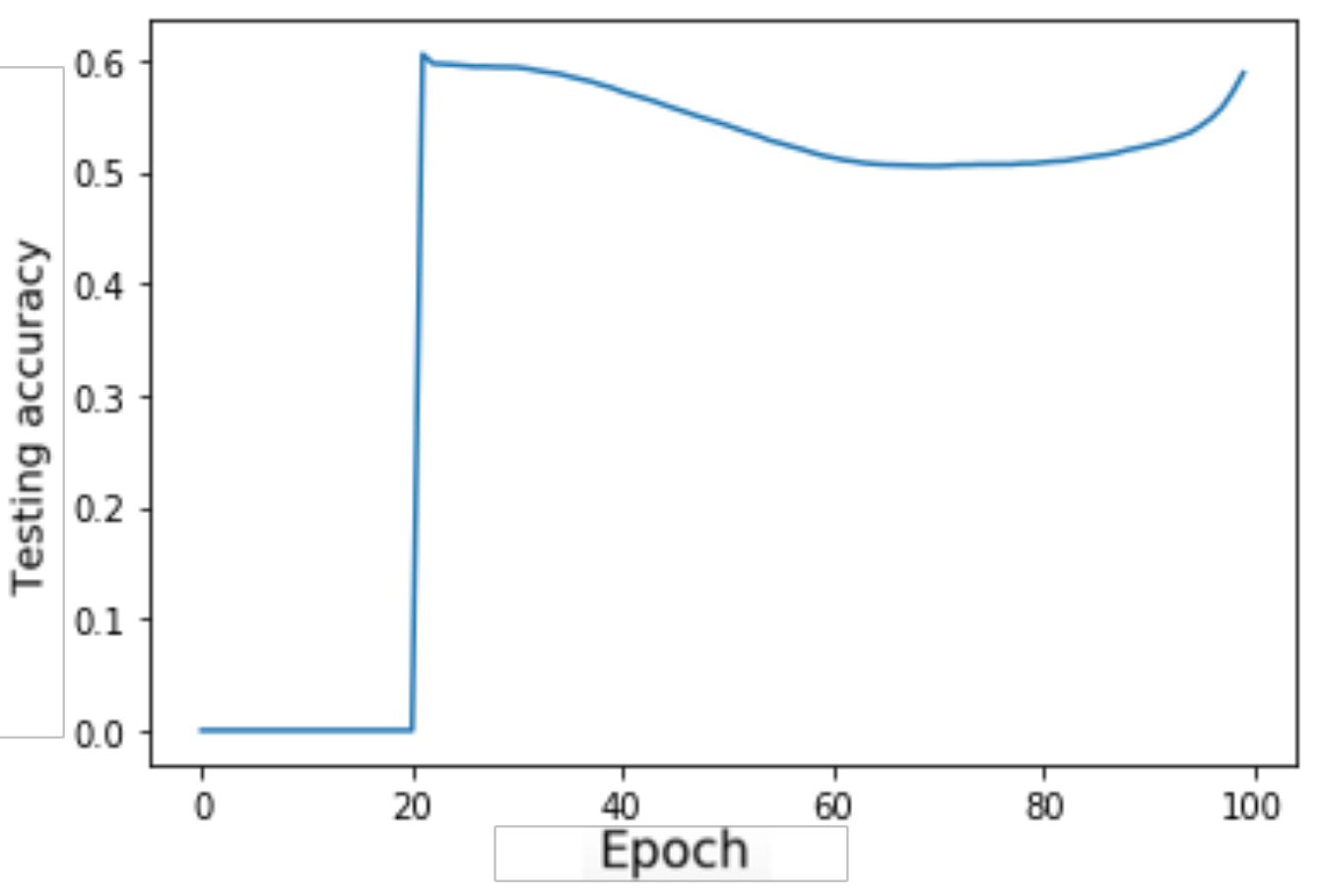}
        \caption{Testing accuracy of single machine training with TensorFlow~v2 code.}
        \label{v2}
    \end{subfigure}
    \centering
    \begin{subfigure}[b]{0.47\textwidth}
        \centering
        \includegraphics[width=\textwidth]{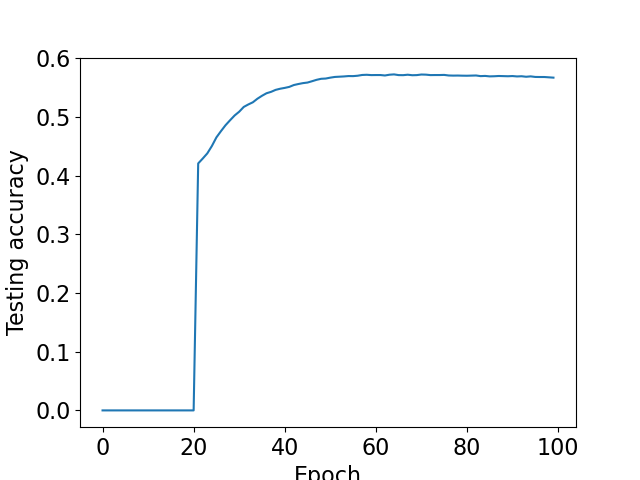}
        \caption{Testing accuracy of distributed training, 8 GPUs.}
        \label{512}
    \end{subfigure}
    \caption{Testing accuracies. The first 20 epochs were pre-training without any clustering, so the clustering accuracy is set to 0.}
    \label{fig-1}
\end{figure}

\begin{figure}
    \centering
    \includegraphics[width=0.7\textwidth]{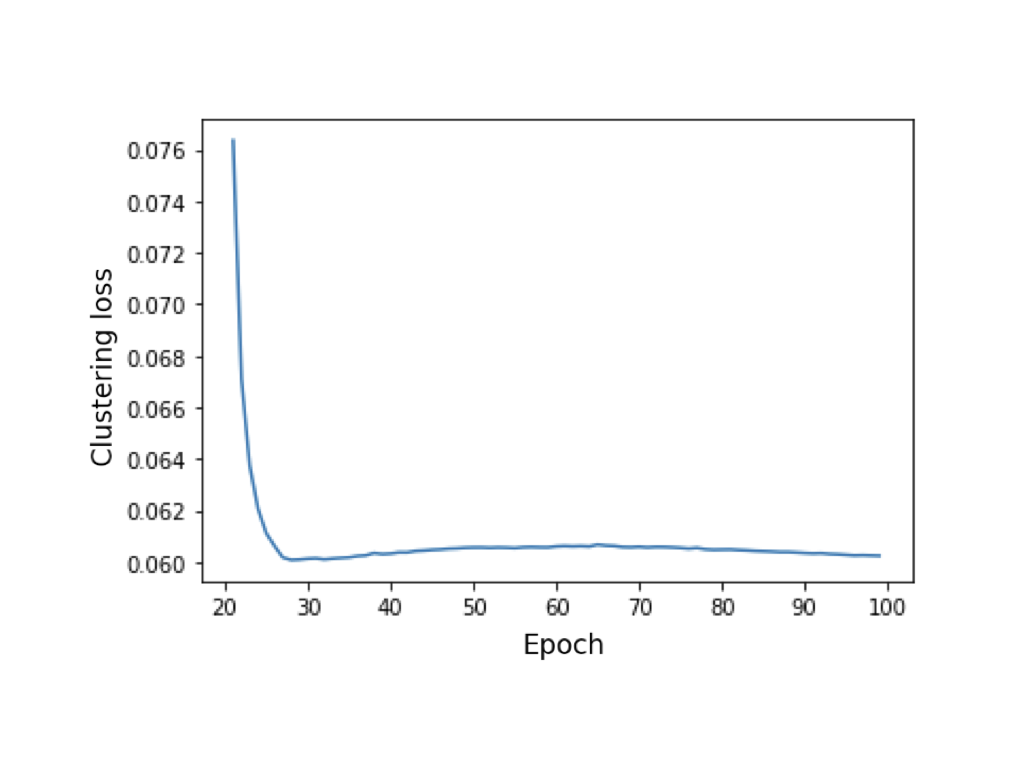}
    \caption{Clustering loss during training of the distributed model on 4 GPUs.}
    \label{loss}
\end{figure}

The model was trained with the same hyper-parameters as found in the original paper, summarised in Table~\ref{tab-2}. It was first pre-trained for 20 epochs with only the classification net, before being trained end-to-end (with the combined classification and clustering loss) for a total of 100 epochs. The training was done on GPU clusters with either 2, 4 or 8 NVIDIA T4 GPUs, each with 16 GB of memory, running Apache Spark 3.0.1 (Amazon EC2 G4dn.xlarge instance). The training time per epoch for different number of GPUs can be seen in Table~\ref{tab-3}, compared to the training time for the single machine codes. The training time can be seen to inversely scale with the number of GPUs. Each of the models, the distributed code as well as the single machine codes, has been trained several times, and the testing accuracy during training has been recorded. A representative plot for each model can be seen in Figure~\ref{fig-1}. The models display very similar training behaviour, which is to be expected if the distributed training works as it should: For all models, the accuracy lies in the interval 0.5 to 0.6 throughout the training, usually increasing slightly with training time and occasionally decreasing over a number of epochs before returning to the stable behaviour. It can be noted that the testing accuracy is far from the $81\%$ validation accuracy obtained in the original paper for all models, including the original one. Since there is virtually no improvement with training, the optimisation algorithms might have found a local minimum. Figure~\ref{loss} shows the clustering loss during training of the distributed model on 4 GPUs. Here we can see a minimum at around epoch 28, and then a slight increase until epoch 65. Some hyper-parameter optimisation was done, testing $\alpha$ (inverse temperature) values that changed by a factor of (1,2,5,10) every epoch following the starting value of 1, changing the batch size to 512, changing the proportionality constant $\beta$ between the classification loss and the clustering loss to ($10^{-1}$,$1$,$10$,$100$) and lowering the learning rate. All trials gave worse or comparable results to those in Table~\ref{tab-2}. Furthermore, using the same hyper-parameters consistently shows the same behaviour as displayed in Figure~\ref{fig-1} across a number of trials, independent on the number of GPUs used for training.

\begin{table}
\centering
\caption{Hyper-parameters used in training.}
\label{tab-2}       
\begin{tabular}{lll}
\hline
Hyper-parameter & Value  \\\hline
Batch size & 1024 \\
Inverse temperature $\alpha$ & Starting at 1, increasing linearly by 2\\
&every following epoch \\
Proportionality constant between\\ classification loss and clustering loss $\beta$ & 10\\ 
Focal loss hyperparameter $\gamma$ & 2\\
Learning rate & Starting at 0.001 and decreases by 2\\ 
&every three epochs until it has reached $10^{-5}$.\\
&Multiplied by a factor equal to the number of GPUs.\\
Optimizer & Adam\\\hline
\end{tabular}
\end{table}
\begin{table}
\centering
\caption{Training times for the single machine codes and the distributed code.}
\label{tab-3}       
\begin{tabular}{lll}
\hline
Code & Training time per epoch  \\\hline
Original code & 4 minutes\\
TensorFlow 2 version of original code& 4 minutes\\
Distributed training, 2 GPUs & 2 minutes\\
Distributed training, 4 GPUs & 1 minute\\
Distributed training, 8 GPUs & 30 seconds\\\hline
\end{tabular}
\end{table}

\section{Next steps}
To make the model fully scalable, we need both the data ingestion and the training to be scalable. We will accomplish this by loading data directly into the distributed file system, bypassing the memory limitations of the main memory, as well as copying only subsets of data onto each GPU. For the data loading, we will use file formats designed for distributed processing of large data sets such as the open-source column-oriented data storage format Parquet or the to high energy physicists well-known ROOT format. For the data distribution, we will compare the Horovod framework to the Maggy~\cite{maggy} framework to see if there are already tools implemented that can be used for this purpose.

We are also actively looking into how the accuracy can be improved to that which was shown in the original paper. Since the optimisation might be stuck in a local minimum, a natural path forward would be to investigate the inverse temperature and potential energy surface defined by the optimisation problem. If the weights of the fully trained, accurate model can be made available to us, this could provide valuable insight without having to look into the inverse temperature. We could initialise our weights with the fully trained parameters, or slight perturbations there of, and see if our training reaches the same parameters again.

\section{Conclusion and outlook}
We have implemented distributed training for the UCluster method and are in the process of making it scalable to any input data size. We migrated the UCluster model to TensorFlow~v2 and added distributed training using the HorovodRunner. After this, the training behaviour of the model is very similar to that of the original code, and this behaviour is consistent over a large number of trials. However, we see a significantly lower accuracy than that reported in the original paper. We are in the process of making the setup fully scalable, as well as troubleshooting the lowered accuracy. The UCluster method is a very general method in the sense that it can be modified for any task in which unsupervised clustering of particles could be used. The generality of the original model together with the scalability added in this project has the potential to be very powerful in processing large amounts of data for a wide variety of tasks at the LHC. The distributed model can already be used as is for very fast training with HDF5 data, a format commonly used in particle physics machine-learning challenges open to researchers outside of the big LHC collaborations. Once it has been made fully scalable, it will be able to train directly on both experimental data and simulated data from the LHC -- possibly requiring some pre-processing of the files. 

\section{Acknowledgements}
This research was partially supported by the Wallenberg AI, Autonomous
Systems and Software Program (WASP) funded by Knut and Alice Wallenberg Foundation, the Center for Interdisciplinary Mathematics at Uppsala University and Combient Competence Centre for Data Engineering Sciences at Uppsala University. 
This project originated in the Scalable Data Science and Distributed Machine Learning course of the WASP graduate school
that was supported by Databricks University Alliance with an infrastructure grant from AWS. We would also like to thank Vinicius Mikuni for help in getting the model set up and Tilo Wiklund for his support as an industrial mentor sponsored by Combient Mix AB, Stocholm.




\bibliography{bib.bib}

\begin{thebibliography}{18}

\bibitem{rev}
D.~Bourilkov, International Journal of Modern Physics A \textbf{34}, 1930019
  (2019)

\bibitem{elias}
K.~Albertsson, P.~Altoe, D.~Anderson, J.~Anderson, M.~Andrews, J.P.A. Espinosa,
  A.~Aurisano, L.~Basara, A.~Bevan, W.~Bhimji et~al., \emph{Machine learning in
  high energy physics community white paper} (2019), \texttt{1807.02876}

\bibitem{hepmllivingreview}
{HEP ML Community}, \emph{{A Living Review of Machine Learning for Particle
  Physics}}, \urlstyle{tt}\url{https://iml-wg.github.io/HEPML-LivingReview/}

\bibitem{Mikuni:2020qds}
V.~Mikuni, F.~Canelli (2020), \texttt{2010.07106}

\bibitem{multiclassdata}
M.~Pierini, J.M. Duarte, N.~Tran, M.~Freytsis, \emph{Hls4ml lhc jet dataset
  (100 particles)} (2020),
  \urlstyle{tt}\url{https://doi.org/10.5281/zenodo.3602254}

\bibitem{gregor_kasieczka_2019_2629073}
G.~Kasieczka, B.~Nachman, D.~Shih, \emph{{R\&D Dataset for LHC Olympics 2020
  Anomaly Detection Challenge}} (2019),
  \urlstyle{tt}\url{https://doi.org/10.5281/zenodo.2629073}

\bibitem{mining}
O.~Cerri, T.Q. Nguyen, M.~Pierini, M.~Spiropulu, J.R. Vlimant, Journal of High
  Energy Physics \textbf{2019} (2019)

\bibitem{spark}
M.~Zaharia, R.S. Xin, P.~Wendell, T.~Das, M.~Armbrust, A.~Dave, X.~Meng,
  J.~Rosen, S.~Venkataraman, M.J. Franklin et~al., Commun. ACM \textbf{59},
  56–65 (2016)

\bibitem{databricks}
\emph{Databricks}, \url{https://databricks.com/} (2021), accessed: 2021-02-28

\bibitem{horovod}
A.~Sergeev, M.D. Balso, arXiv preprint arXiv:1802.05799  (2018)

\bibitem{ABC}
V.~Mikuni, F.~Canelli, Eur. Phys. J. Plus \textbf{135}, 463 (2020),
  \texttt{2001.05311}

\bibitem{deepk}
M.M. Fard, T.~Thonet, E.~Gaussier, \emph{Deep $k$-means: Jointly clustering
  with $k$-means and learning representations} (2018), \texttt{1806.10069}

\bibitem{tensorflow}
M.~Abadi, A.~Agarwal, P.~Barham, E.~Brevdo, Z.~Chen, C.~Citro, G.S. Corrado,
  A.~Davis, J.~Dean, M.~Devin et~al., \emph{{TensorFlow}: Large-scale machine
  learning on heterogeneous systems} (2015), software available from
  tensorflow.org, \urlstyle{tt}\url{https://www.tensorflow.org/}

\bibitem{antikt}
M.~Cacciari, G.P. Salam, G.~Soyez, JHEP \textbf{04}, 063 (2008),
  \texttt{0802.1189}

\bibitem{PDG}
M.~Tanabashi et~al. (Particle Data Group), Phys. Rev. D \textbf{98}, 030001
  (2018)

\bibitem{hdf5}
{The HDF Group}, \emph{{Hierarchical data format version 5}} (2000-2010),
  \urlstyle{tt}\url{http://www.hdfgroup.org/HDF5}

\bibitem{uclustergit}
V.~Mikuni, \emph{Ucluster}, \url{https://github.com/ViniciusMikuni/UCluster}
  (2020)

\bibitem{maggy}
M.~Meister, S.~Sheikholeslami, A.H. Payberah, V.~Vlassov, J.~Dowling,
  \emph{Maggy: Scalable Asynchronous Parallel Hyperparameter Search}, in
  \emph{Proceedings of the 1st Workshop on Distributed Machine Learning}
  (Association for Computing Machinery, New York, NY, USA, 2020),
  DistributedML'20, p. 28–33, ISBN 9781450381826,
  \urlstyle{tt}\url{https://doi.org/10.1145/3426745.3431338}

\end{thebibliography}
%
%
%
%

\end{document}